\documentclass{webofc}
\usepackage[varg]{txfonts}   
\usepackage{cleveref}
\usepackage{mathtools}
\usepackage{xcolor}
\begin{document}
\title{Kernel controlled real-time Complex Langevin simulation}
%
%

\author{\firstname{Daniel} \lastname{Alvestad}\inst{1}\fnsep\thanks{\email{daniel.alvestad@uis.no}}
\and
       \firstname{Rasmus} \lastname{Larsen}\inst{1}\fnsep\thanks{\email{rasmus.n.larsen@uis.no}}
        \and
       \firstname{Alexander} \lastname{Rothkopf}\inst{1}\fnsep\thanks{\email{alexander.rothkopf@uis.no}}
}

\institute{Faculty of Science and Technology, University of Stavanger, NO-4021 Stavanger, Norway}

\abstract{%
This study explores the utility of a kernel in complex Langevin simulations of quantum real-time dynamics on the Schwinger-Keldysh contour. We give several examples where we use a systematic scheme to find kernels that restore correct convergence of complex Langevin. The schemes combine prior information we know about the system and the correctness of convergence of complex Langevin to construct a kernel. This allows us to simulate up to $1.5\beta$ on the real-time Schwinger-Keldysh contour with the $0+1$ dimensional anharmonic oscillator using $m=1,\lambda=24$, which was previously unattainable using the complex Langevin equation.
}
\maketitle
\section{Introduction}

The notorious sign problem \cite{gattringer_approaches_2016,Pan:2022fgf} prevents the computation of the real-time dynamics of strongly correlated quantum fields by means of established Monte-Carlo methods. It thus constitutes a major roadblock in the study of phenomenologically relevant transport phenomena in various field of physics, be it in the quark-gluon plasma created in relativistic heavy-ion collisions or in the functional materials studied in condensed matter systems. 

One promising strategy to enable direct simulations of the real-time dynamics is the complex Langevin approach \cite{namiki_stochastic_1992}, based on stochastic quantization \cite{damgaard_stochastic_1987} (for other approaches see e.g. \cite{berger_complex_2021}) of Feynman's path integral $\langle O\rangle = \int {\cal D}\phi\, O\, {\rm exp}[iS[\phi]]$. It proposes to complexify the degrees of freedom of the system $\phi\to\phi^R+i\phi^I$ and evolve them stochastically in an artificial additional time, so called Langevin time $\tau_L$ with Gaussian noise $\eta$ 
\begin{align}
\frac{d\phi^R}{d\tau_L}={\rm Re}\left[ \left. i\frac{\delta S[\phi]}{\delta \phi}\right|_{\phi=\phi^R+i\phi^I}\right]+\eta(\tau_L),\quad \frac{d\phi^I}{d\tau_L}={\rm Im}\left[ \left. i\frac{\delta S[\phi]}{\delta \phi}\right|_{\phi=\phi^R+i\phi^I}\right].
\end{align}
Instead of describing the individual realizations of the complexified system with a Langevin equation we can consider the evolution of the distribution of the d.o.f. via the associated Fokker-Planck equations. There exists two of those, a genuine FP equation, which describes the real distribution $P[\phi^R,\phi^I]$ of the complexified $\phi$ and a complex FP equation, which describes the distribution $\rho(\phi)$ and whose late time limit must be $\lim_{\tau_L\to\infty}\rho(\phi)\propto{\rm exp}[iS]$ for complex Langevin to succeed.

This condition is one among several (for the boundary criterion see \cite{Aarts:2011ax}), which over the past years have been explored to verify whether complex Langevin dynamics converge to the correct solution. The possibility to converge to incorrect solutions, together with the occurrence of so-called runaway trajectories constitute the two main drawbacks of this approach (see e.g. \cite{Seiler:2017wvd}).

In a previous publication \cite{Alvestad:2021hsi} we have shown how to avoid runaway trajectories, by introducing inherently stable solvers to the solution of the complex Langevin dynamics. In contrast to the adaptive step size prescription \cite{AartsJames2010} discussed in the literature, the implicit solvers also provide a novel mechanism to regularize the underlying path integral we wish to simulate. The simplest extension of the often deployed forward Euler scheme is the generalized Euler-Maruyama (EM) scheme. It evolves each discretized field $\phi_j$ via time step $\varepsilon$ according to
\begin{align}
\phi_j^{n+1}=\phi_j^{n}+i\varepsilon\Big[\theta \frac{\delta S^{n+1}}{\delta \phi_j}+(1-\theta)\frac{\delta S^{n+1}}{\delta \phi_j}\Big]+\sqrt{\varepsilon}\eta_j^{n},
\end{align} 
where the implicitness parameter $\theta$ governs the stability properties. For $\theta\geq 1/2$ the approach is unconditionally stable. The stability and inherent ability of this solver to regularize the path integral, allowed us in ref.~\cite{Alvestad:2021hsi}  to simulate, for the first time, the real-time dynamics of the strongly coupled anharmonic oscillator on the canonical Schwinger-Keldysh (SK) contour without tilt. 

The real-time extent of these simulations was very small $t_{\rm max}m=0.5$ in order for the naive complex Langevin equations to converge to the correct solution. Of course to obtain phenomenologically relevant insight these simulations must be extended to larger real-times, which is the aim of the study presented in this contribution.

As one extends the real-time extent of the Schwinger-Keldysh contour it is known that naive complex Langevin fails to converge correctly \cite{BergesSexty2007}. This phenomenon has been investigated in details in the community and it was found that the appearance of so called boundary terms \cite{Aarts:2011ax} signals the failure of convergence. In the context of complex Langevin for gauge theories the gauge cooling approach \cite{Seiler2013} and more recently dynamical stabilization \cite{Aarts:2016qhx} have been proposed as means to avoid boundary terms and have led to impressive improvements in the simulations of QCD at finite Baryo-chemical potential \cite{Attanasio:2022mjd}. The combination of the two techniques is also deployed to study the real-time dynamics of gauge fields \cite{Boguslavski:2022vjz}.

One way to tame the sign problem and the associated incorrect convergence of CL in the past was to construct coordinate transformations or field redefinitions \cite{Aarts:2012ft}. They can all be summarized by a general modification of the complex Langevin equations in terms of a so called \textit{kernel}. The role of kernels for the convergence of CL has been explored early on in simple models in refs.~\cite{Okamoto:1988ru}. 

In our study we focus on kernelled complex Langevin and will introduce a novel strategy based on prior knowledge that allows us to learn optimal kernels to restore correct convergence. Note that the sign problem is known to be NP-hard \cite{Troyer:2004ge} and thus no generic solution strategy is believed to exist. In order to make progress we thus believe system-specific information needs to be incorporated into the CL simulations and our learning strategy offers exactly that.

\section{Kernelled Complex Langevin}

\begin{figure}
\centering
\includegraphics[scale=0.15]{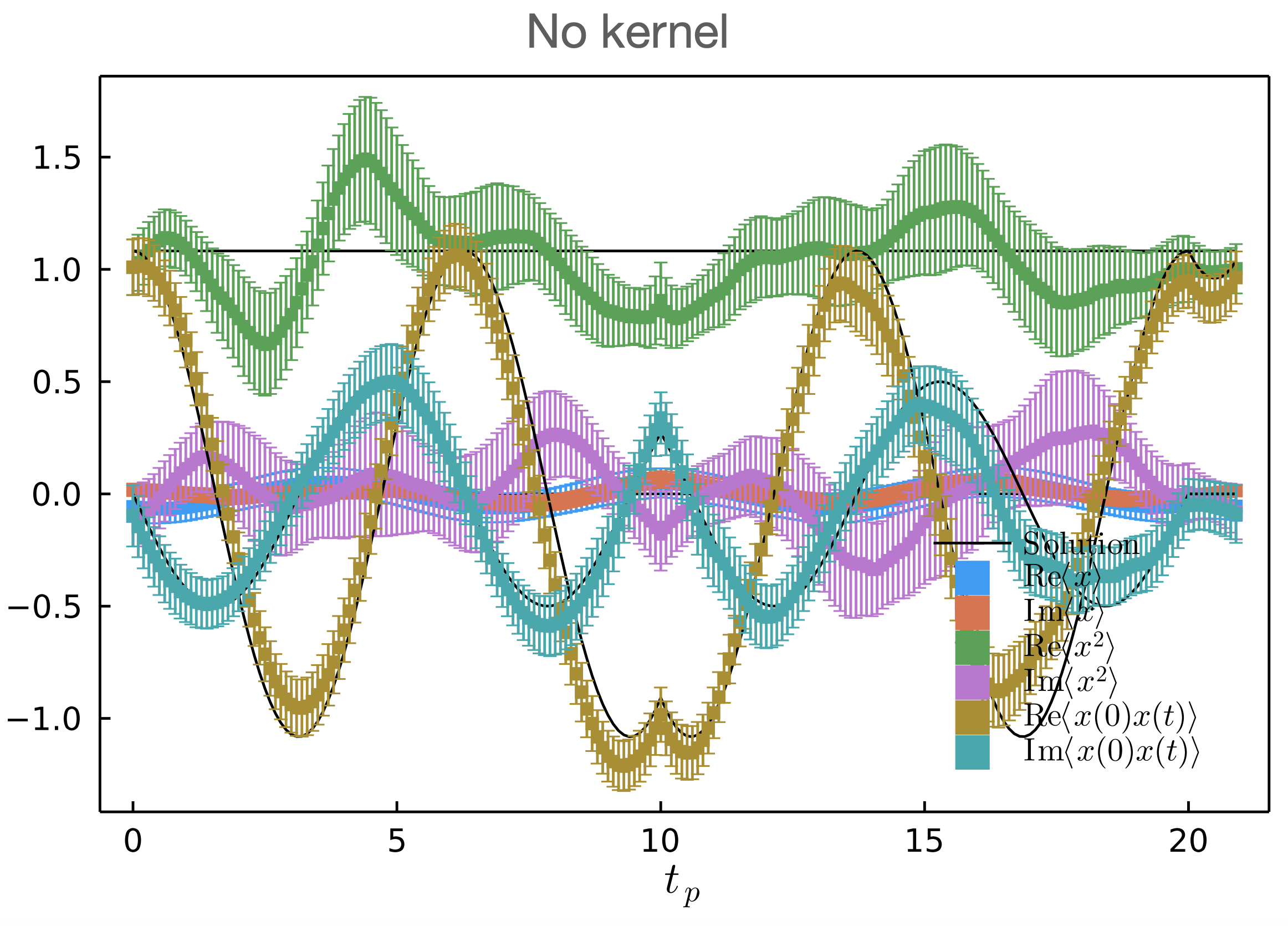}
\includegraphics[scale=0.15]{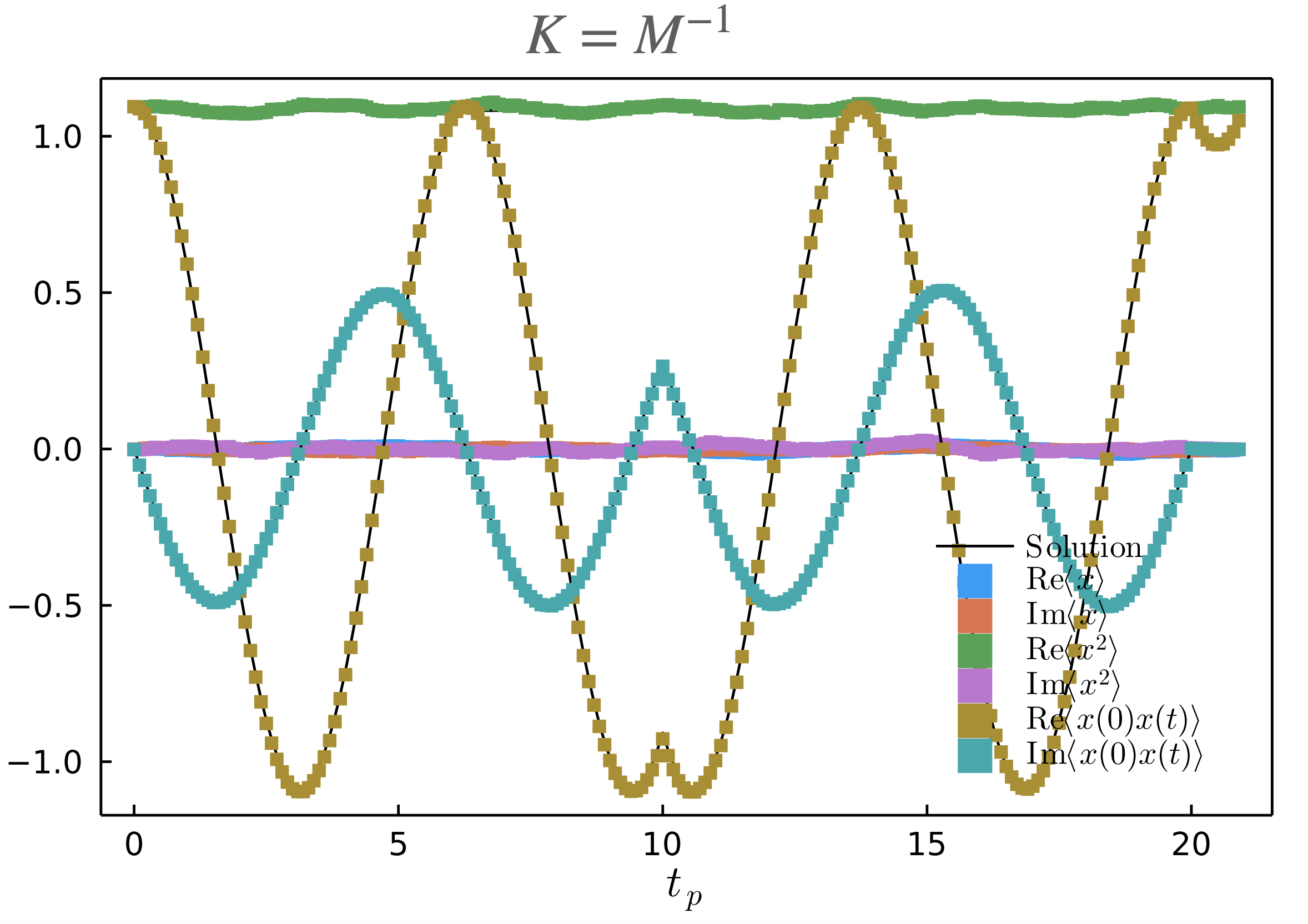}
\caption{(left) The CL dynamics of the harmonic oscillator on the SK contour with real-time extent $m t_{\rm max}=10$ in the absence of a kernel. (right) Same system after introducing the manually constructed kernel of \cref{eq:ManualK}.}\label{fig:FreeTheory}
\end{figure}

It is known \cite{namiki_stochastic_1992} that real-valued Langevin equations offer the freedom to introduce a real-valued kernel in the stochastic dynamics ($dW$ denotes a Wiener process)
\begin{align}
d\phi=-K\frac{\delta S}{\delta \phi}d\tau_L+\frac{\delta K}{\delta \phi}d\tau_L+\sqrt{K}dW.
\end{align}
It will change the approach to the unique stationary distribution but does not change its form. In complex Langevin we may introduce a complex valued kernel (making sure that it factorizes $K=H^\dagger H$ so that one can take the square root), which amounts to a non-neutral modification of the dynamics
\begin{align}
&\frac{d\phi^R}{d\tau_L}={\rm Re}\left[ iK[\phi]\frac{\delta S[\phi]}{\delta \phi} + \frac{\delta K}{\delta \phi} \right]_{\phi=\phi^R+i\phi^I}+{\rm Re}[H[\phi^R+i\phi^I]]\eta(\tau_L),\\
&\frac{d\phi^I}{d\tau_L}={\rm Im}\left[ iK[\phi]\frac{\delta S[\phi]}{\delta \phi}+\frac{\delta K}{\delta \phi}\right]_{\phi=\phi^R+i\phi^I}+{\rm Im}[H[\phi^R+i\phi^I]]\eta(\tau_L).
\end{align}
Through a tuning of such a kernel it is thus fathomable that the correct late time distribution can be recovered. One example where a manually constructed kernel can be used to restore correct convergence is in the harmonic oscillator on the SK contour. It was shown in ref.~\cite{Okamoto:1988ru} that by turning the drift term of the CL equation into $-\phi$ through the application of a kernel, correct convergence can be achieved in a simple one d.o.f. model. We now understand this result from connecting complex Langevin to the thimble structure of the theory (see section \cref{sec:Thimbles}). The construction for the harmonic oscillator proceeds similarly
\begin{align}
S=\int_{\rm SK} dt\, {\bf \phi}(t)M\phi(t), \quad \to\quad K=iM^{-1} \quad\to\quad K\delta S/\delta \phi=-\phi.\label{eq:ManualK}
\end{align}
Using the implicit EM scheme with $\theta=0.6$ at $\beta=1$ and collecting statistics over hundred different trajectories up to $m\tau_L=100$, we show in \cref{fig:FreeTheory} how the introduction of the manual kernel of \cref{eq:ManualK} (right) tames the large noise and incorrect convergence of the naive CL dynamics (left). The colored data points show the real- and imaginary part of different correlators evaluated along the SK contour. For $mt_p\leq mt_{\rm max}=10$ the quantities are evaluated on the forward contour, for $10\leq mt_p\leq 20$ on the backward contour and for the remaining values of $t_P$ on the Euclidean branch (blue and orange: $\langle x\rangle$, green and pink $\langle x^2\rangle$ and brown and turquoise $\langle x(0)x(t_P)\rangle$). The solid lines indicate the known values obtained from the Schr\"odinger equation.

\begin{figure}
\centering
\includegraphics[scale=0.15]{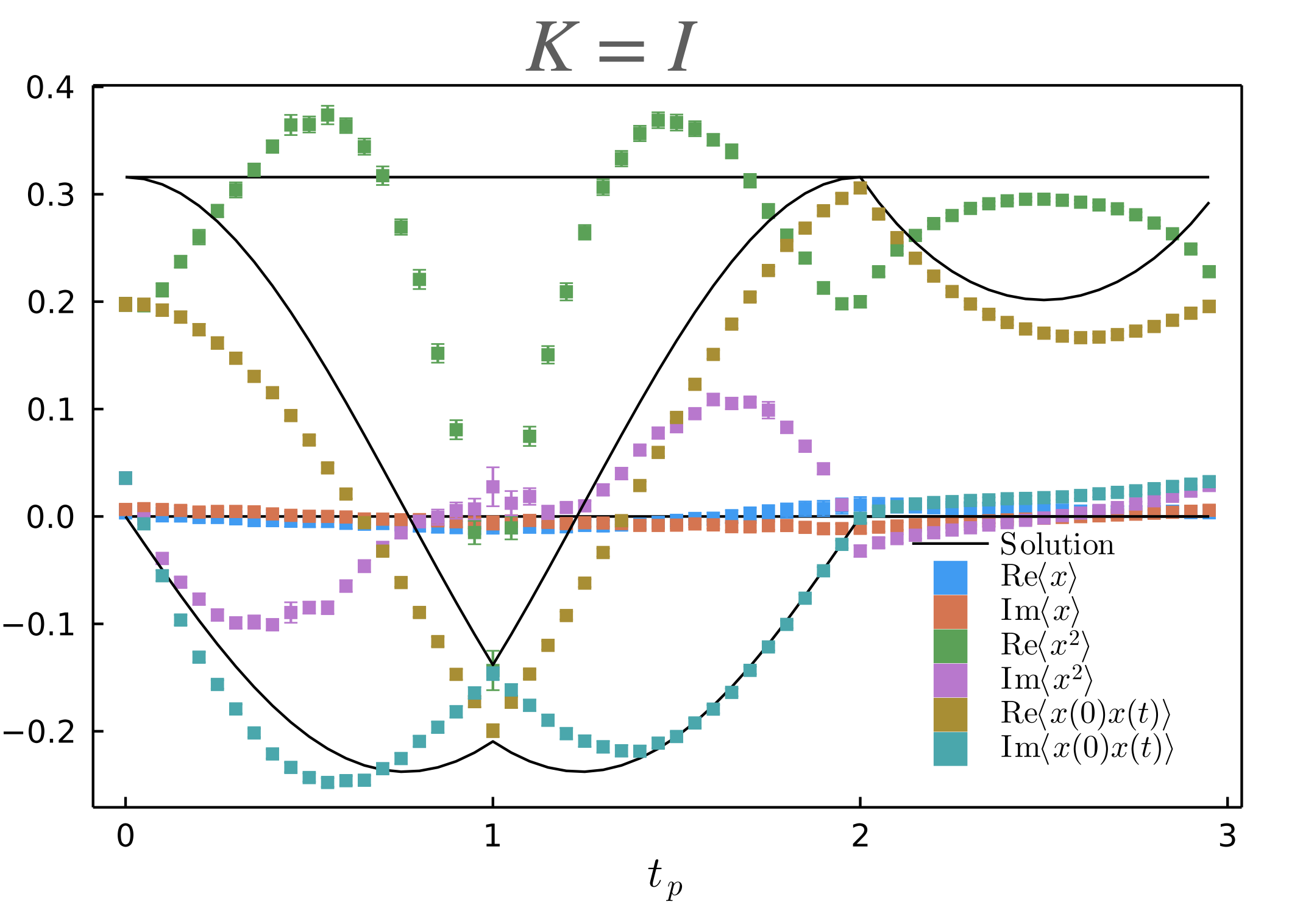}
\includegraphics[scale=0.135]{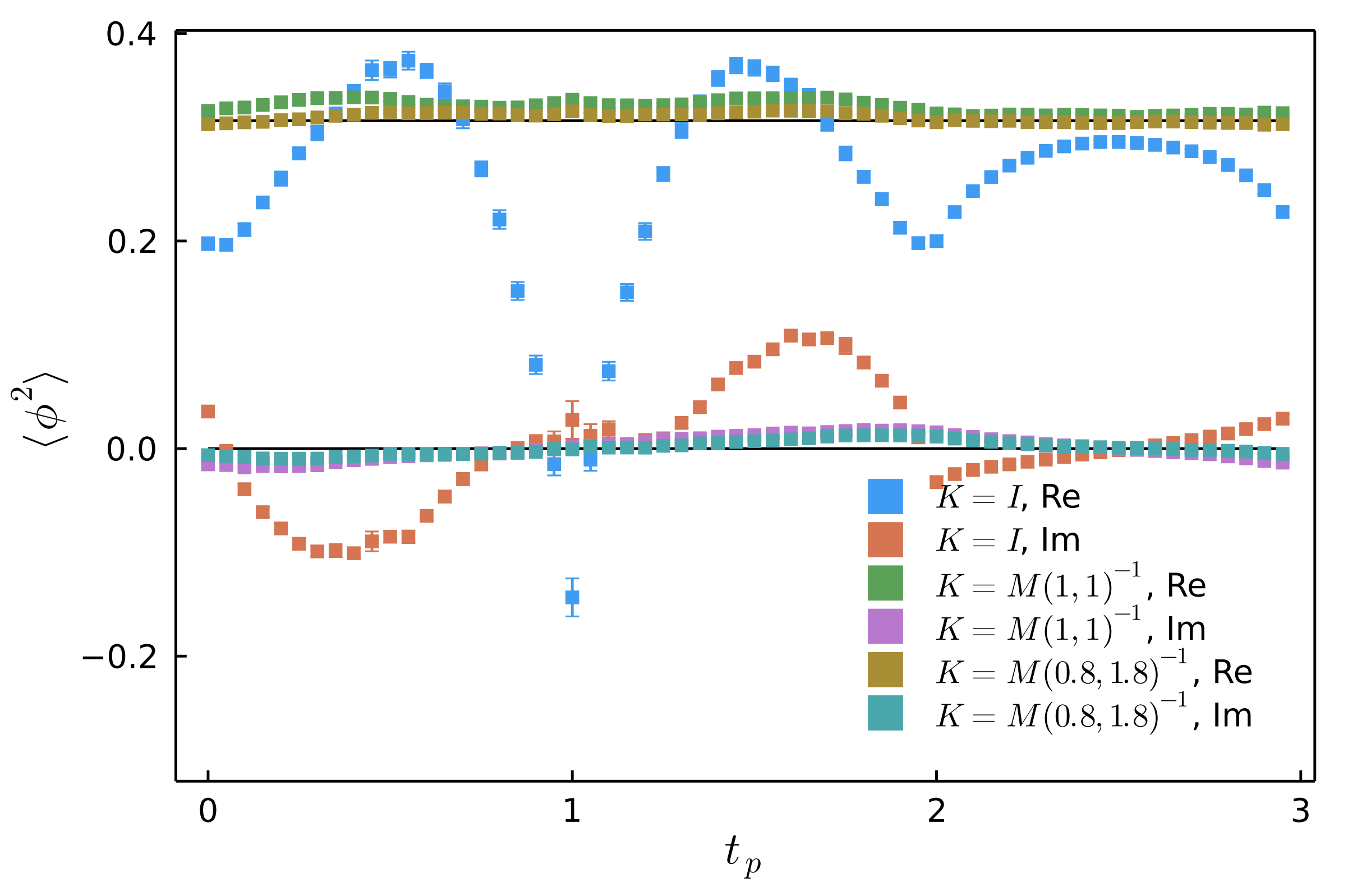}
\caption{(left) The CL dynamics of the strongly coupled anharmonic oscillator on the SK contour with real-time extent $m t_{\rm max}=1$ in the absence of a kernel. Note the incorrect convergence in all observables. (right) Same system after introducing the manually constructed kernel of \cref{eq:ManualK}.}\label{fig:IntTheoryFreeK}
\end{figure}

Interestingly the kernel manually constructed in the free theory also allows us to improve convergence for the strongly coupled anharmonic oscillator with $\lambda=24$ a common benchmark system in the literature. Deploying $K$ of \cref{eq:ManualK} naively allows us to extend correct convergence up to $mt_{\rm max}=0.75$. Modifying the relative weight between the kinetic and potential term in the kernel matrix $M$ by multiplying the derivative term by $g=0.8$ and the potential term by $m_g=1.8$, values that were heuristically found, we even manage to get to $mt_{\rm max}=1$. In the right panel of \cref{fig:IntTheoryFreeK} we plot the values of $\langle x^2\rangle$ for different choices of kernel and in the left panel show the anharmonic oscillator results wihthout the presence of a kernel.

These results are encouraging as they show that a well chosen kernel indeed can help restore correct convergence. The central challenge however lies in the fact that no systematic prescription exists on how to modify the kernel further to restore convergence at later real-time, i.e. for a larger extent of the SK contour.

\section{Learning optimal kernels for correct convergence}

Our proposal therefore is to take inspiration from the machine learning community and attempt to systematically find optimal kernels based on prior knowledge available on the system. Such prior information comes the form of e.g. the symmetries of the system, Euclidean correlators accessible in conventional simulations and also in the form of the correctness criterion that requires that boundary terms are absent. For each type of information we propose to construct a cost functional $L^{\rm sym}$,$L^{\rm Eucl}$ and $L^{\rm BT}$ whose value taken together $L^{\rm prior}=L^{\rm sym}+L^{\rm Eucl}+L^{\rm BT}$ shall be minimized by tuning the entries of the kernel matrix $K$. As a first step we will consider only field-independent kernels for simplicity, which can be seen as the lowest order in an expansion in basis functions depending on the field $\phi$. In this study we will only include $L^{\rm sym}$ and $L^{\rm Eucl}$, where we exploit that the two-point functions are known in Euclidean time and that time-translation invariance enforces that the equal-time correlation functions must remain constant over the whole of the SK contour.

To minimize the loss we have to compute gradients of the cost functionals and in turn evaluate the dependence of observables on the values of the kernel $K$. This entails taking the derivative w.r.t. the whole of the CL simulation. Such operation in principle can be accomplished by the use of modern autodifferentiation techniques \cite{baydin2018automatic}, but we have found that due to the stiffness of the CL equation in real-time many of these direct methods in their standard implementation are too costly in practice. One path forward in the future is to implement the adjoint method specific to real-time CL.

Instead we constructed a heuristic low-cost gradient with which we manged to significantly reduce the values of the true cost functional $L^{\rm prior}$. This low-cost gradient is obtained from approximating the true gradient of the functional $L_{D} =   \left\langle \Big| D(x) \cdot (-x) - ||D(x)||\;||x|| \Big|^\xi\right\rangle$, with $D=K\partial S/\partial\phi$, which is designed to favor the absence of boundary terms by pulling the field degrees towards the origin. Best results were obtained when choosing $1<\xi<2$.  The efficient evaluation of the low-cost gradient from $L_D$ was achieved by using autodifferentiation.

\begin{figure}
\centering
\includegraphics[scale=0.15]{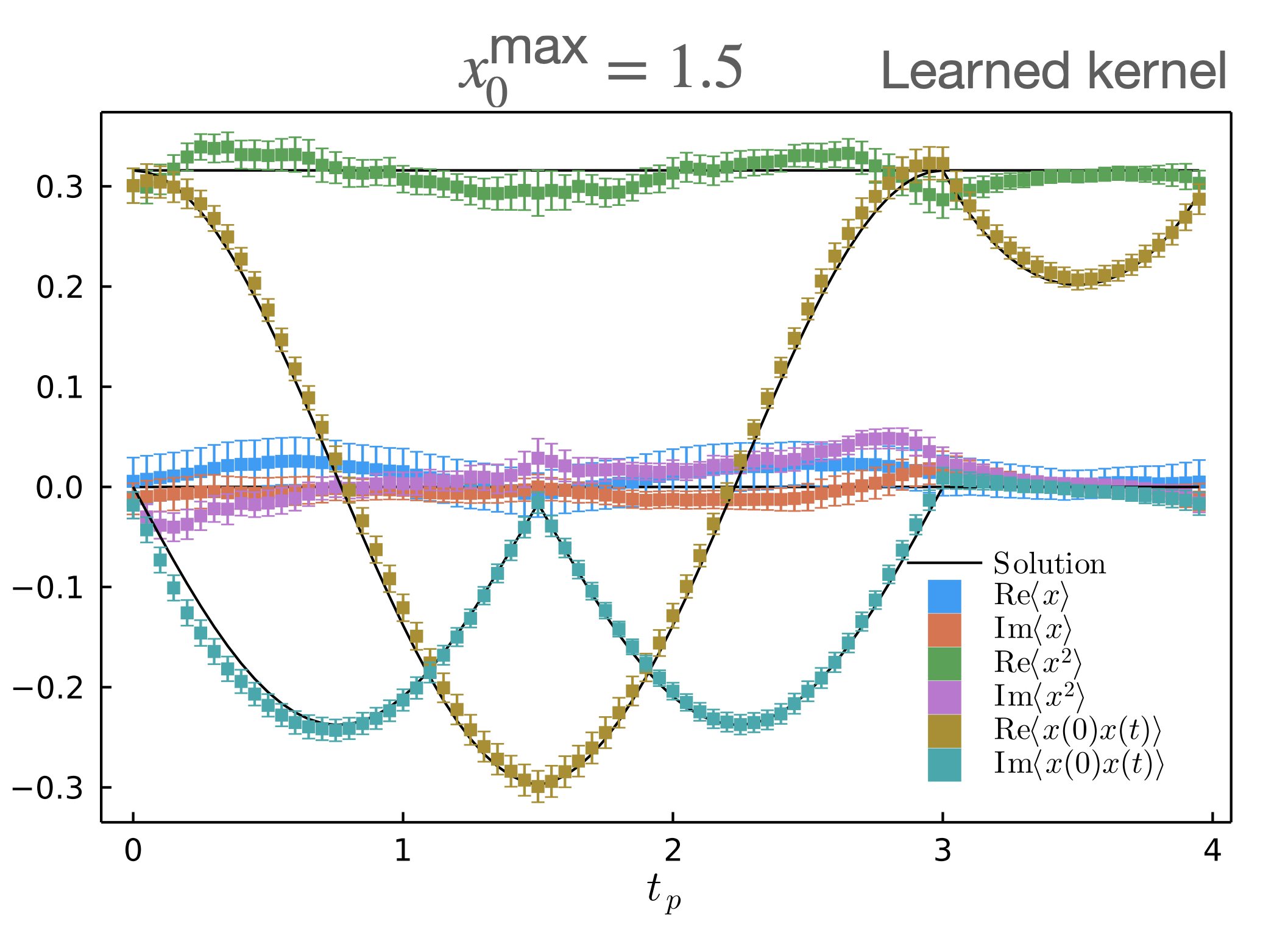}
\includegraphics[scale=0.15]{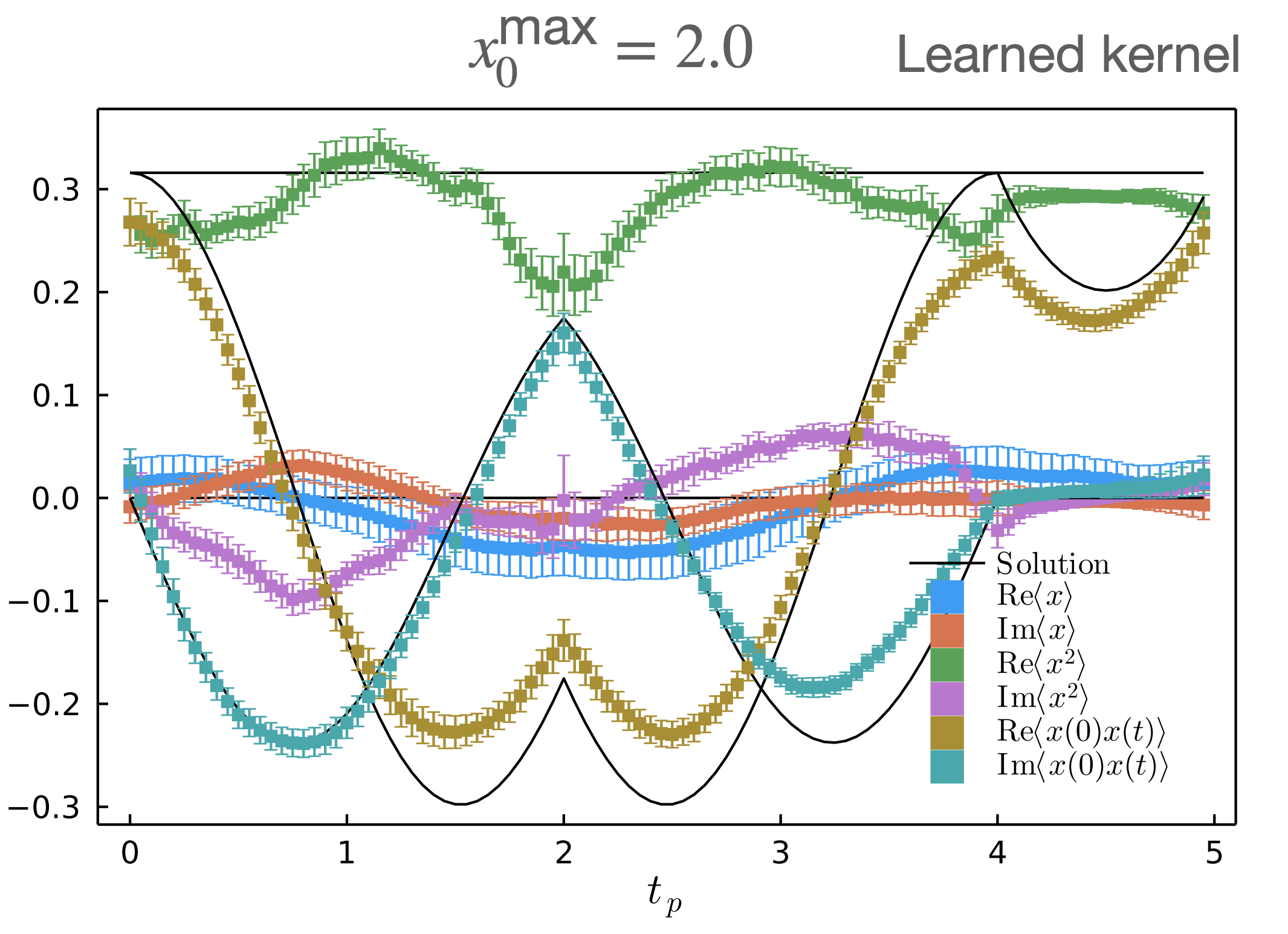}
\caption{(left) CL dynamics of the anharmonic oscillator in the presence of our learned field-independent optimal kernel at $mt_{\rm max}=1.5$. Correct convergence is restored (c.f. left panel of \cref{fig:IntTheoryFreeK}) (right) Results of the CL simulations at $mt_{\rm max}=2$, where we fail to learn a field-independent kernel to restore correct convergence.}\label{fig:IntTheoryLearnedK}
\end{figure}

By iteratively finding the optimal values for the kernel using the low-cost gradient and monitoring the success of the optimization via $L^{\rm prior}$ we managed to extend correct convergence of CL for the strongly coupled anharmonic oscillator up to $mt_{\rm max}=1.5$ as shown in the left panel of \cref{fig:IntTheoryLearnedK}. This constitutes a threefold improvement over the previously achieved record for the real-time simulation of the anharmonic oscillator in complex Langevin.

Our approach of course has its limitations. Using a field independent kernel, we fail to restore convergence when extending the SK contour to $mt_{\rm max}=2$, as shown on the right of \cref{fig:IntTheoryLearnedK}. There are several possible reasons for this failure: the use of the low-cost gradient may simply fail to locate the true minimum of $L^{\rm prior}$.  This is related to the fact that from simple models it is known that pulling the complexified fields towards the origin has a beneficial effect if the theory has its critical points at the origin. In general the critical points of the theory may lie away from the origin, rendering the idea behind the gradient based on $L_D$ moot. We will most probably require a field dependent kernel to restore correct convergence for larger real-time extent a direction of future work.

\subsection{Connection to Thimbles}
\label{sec:Thimbles}

Let us briefly mention an interesting connection between the introduction of a kernel and the thimble structure of the simulated theory. As was shown in ref.~\cite{Okamoto:1988ru} in the simple complex Gaussian model $S=\frac{1}{2}\phi^2$, the kernel that turns the drift term into the simple form $iK\delta S/\delta\phi=-\phi$ restores correct convergence. We now understand this behavior through the thimble structure of the model. There exists a single thimble oriented downward 45 degrees in the complex plane passing through the origin. While naive complex Langevin samples preferentially parallel to the real $\phi^R$ axis, the kernel rotates the sampling exactly onto the thimble. This restores correct convergence, as sampling along the thimble has a well-defined statistical meaning. 

In systems where multiple thimbles are present rotating the sampling with a field independent kernel only works in certain parameter ranges and in general a field dependent kernel is required to make sure that the physics of different thimbles is captured.

\section{Conclusion}

Bringing together the concept of kernelled complex Langevin with a machine learning inspired strategy to locate their optimal values based on prior information we propose to systematically find kernels that restore correct convergence in simulations of strongly correlated quantum systems on the real-time SK contour. We have shown that the strategy is viable by extending correct convergence for the anharmonic oscillator threefold beyond the previous state-of-the-art. Exploration of field dependent kernels and the implementation of custom autodifferentiation methods, such as adjoint methods for real-time complex Langevin are work in progress.

\section{Acknowledgments}
The team of authors gladly acknowledges supported by the Research Council of Norway under the FRIPRO Young Research Talent grant 286883 and draws on computing resources provided by  
UNINETT Sigma2 - the National Infrastructure for High Performance Computing and Data Storage in Norway under project NN9578K-QCDrtX "Real-time dynamics of nuclear matter under extreme conditions"

\bibliography{references}

\end{document}